\RequirePackage{ifpdf}
\documentclass{PoS}
\pdfoutput = 1
\usepackage{cite}
\usepackage{amsmath}  
\usepackage{amssymb} 
\usepackage{dsfont}
 \usepackage{multirow}
\usepackage{psfrag,scalefnt}
\usepackage{paralist}

\newcommand{\TeV}{\,\text{TeV}}

\newcommand{\muEW}{{\mu_\text{ew}}}

\newcommand{\im}{\text{Im}}
\usepackage{color}
\let\normalcolor\relax

\def\epe{\varepsilon'/\varepsilon}

\newcommand{\gev}{\, {\rm GeV}}

\newcommand{\IM}{{\rm Im}}
\newcommand{\RE}{{\rm Re}}

\newcommand{\be}{\begin{equation}}
\newcommand{\ee}{\end{equation}}
\newcommand{\bea}{\begin{eqnarray}}
\newcommand{\eea}{\end{eqnarray}}

\newcommand{\ba}{\begin{array}}
\newcommand{\ea}{\end{array}}

\newcommand{\ord}{{\cal O}}

\newcommand{\bsi}{B_6^{(1/2)}}
\newcommand{\bei}{B_8^{(3/2)}}

\title{$\varepsilon'/\varepsilon$ in the Standard Model and Beyond: 2021}

\ShortTitle{$\varepsilon'/\varepsilon$ in the  Standard Model and Beyond: 2021}

\author{\speaker{Andrzej J. Buras}%
  \\
TUM-IAS, Lichtenbergstr. 2a, D-85748 Garching, Germany, \\
Technical University Munich, Physics Department, D-85748 Garching, Germany,\\
E-mail: \email{aburas@ph.tum.de}}

\abstract{The present status of the ratio $\varepsilon'/\varepsilon$ in the Standard Model (SM) is summarized. In particular we stress the differences between three approaches
  that attempt to calculate $\varepsilon'/\varepsilon$ within the SM: Lattice QCD, Dual QCD
  and Chiral Perturbation Theory. As presently there is  still a significant
  room left for new physics (NP) contributions, we summarize the present status
  of the technology that has been recently developed with the goal to analyze
  $\varepsilon'/\varepsilon$ and its correlations with other observables within the Weak Effective Theory (WET) and the Standard Model Effective Field Theory (SMEFT). We also make
  a few comments on the interpretation of the main dynamics behind the $\Delta I=1/2$ rule in $K\to\pi\pi$ decays.
 }

\FullConference{CKM Workshop 2021, 22-26 November 2021, Melbourne, Australia
}

\begin{document}

\section{Overture}
The ratio $\epe$ measuring the size of the direct CP violation in $K_L\to\pi\pi$  decays ($\varepsilon^\prime$) relative to the indirect one described by $\varepsilon$ is very sensitive to new sources of CP violation. As such it played a prominent role in particle physics already for 45 years \cite{Buras:2020wyv}.

 Due to the smallness of  $\epe$ its measurement required  heroic efforts  in the 1980s and the 1990s on both sides of the Atlantic  with final results
presented by  NA48 and KTeV collaborations at the beginning of this millennium.
On the other hand, even 45 years after the first calculation of $\epe$ we
do not know to which degree the Standard Model (SM)  agrees with this data and
how large is the room left for new physics (NP) contributions to this ratio.
This is due to significant non-perturbative (hadronic) uncertainties accompanied by  partial cancellations between the QCD penguin  contributions and electroweak penguin contributions. In addition to the calculation of hadronic matrix elements of the relevant operators including isospin breaking effects and QED corrections, it is crucial to evaluate accurately the Wilson coefficients
of the relevant operators. While the significant control over the latter short distance effects has been achieved already in the early 1990s, with several improvements since then, different views on the non-perturbative contributions to $\epe$ have been expressed by different authors over last thirty years. In fact
even  at the dawn of the 2020s the uncertainty in the room left for NP contributions to $\epe$ is still very significant, which I find to be very exciting.

As in addition to \cite{Buras:2020wyv}, this topic is presented
in my recent book \cite{Buras:2020xsm} and in my recent contribution to Kaon 2019 \cite{Buras:2019vik} some overlap with these reviews is unavoidable but I will
attempt to reduce it as far as it is possible without loosing most important information.

Let us then look first at the effective Hamiltonian relevant for $K\to\pi\pi$ decays at the low energy scales. It has the following general structure:
\be\label{BSMH}
{\cal H}_\text{eff}= \sum_i C_i \mathcal{O}_i^\text{SM}+\sum_j C_j^\text{NP} \mathcal{O}_j^\text{NP}\,,\qquad C_i=C_i^\text{SM}+\Delta_i^\text{NP},
\ee
where
\begin{itemize}
\item
$\mathcal{O}_i^\text{SM}$ are local operators present in the SM
and $\mathcal{O}_j^\text{NP}$ are new operators having typically new Dirac structures, in particular scalar-scalar and tensor-tensor ones. 
\item
  $C_i$ and   $C_j^\text{NP}$ are the Wilson coefficients of these operators. NP effects
  modify not only the Wilson coefficients of SM operators but also generate
  new operators with non-vanishing $C_j^\text{NP}$.
\end{itemize}

The amplitude for the transition $K\to\pi\pi$  
can now be written as follows
\be\label{Kpipi}
{\cal A}(K\to \pi\pi)= \sum_i C_i\langle \pi\pi| \mathcal{O}_i^\text{SM}|K\rangle
+\sum_j C_j^\text{NP} \langle \pi\pi| \mathcal{O}_j^\text{NP}|K\rangle\,.
\ee

The coefficients $C_i$ and $C_j^\text{NP}$ can  be calculated in the renormalization group (RG) improved perturbation theory.  The status of these calculations is by now very advanced, as reviewed in 2014 in \cite{Buras:2011we} including also rare meson decays and quark mixing. I plan to update this review this year.
The complete NLO corrections to $K\to\pi\pi$
have been calculated almost 30 years ago \cite{Buras:1991jm,Buras:1992tc, Buras:1992zv, Buras:1993dy, Ciuchini:1992tj, Ciuchini:1993vr}. The dominant NNLO QCD
corrections to electroweak penguin (EWP) contributions have been presented in
\cite{Buras:1999st} and those to QCD penguins (QCDP) in \cite{Gorbahn:2004my,Cerda-Sevilla:2016yzo, Cerda-Sevilla:2018hjk}. These NNLO corrections reducing various scale and renormalization scheme dependences  will play 
 a significant role when the calculation of hadronic matrix elements will be brought under control.  On the whole, the status of present short distance (SD) contributions to $\epe$ is satisfactory.

The evaluation of the hadronic matrix elements is a different story. In
$K\to\pi\pi$ decays, we have presently three approaches  to our disposal:
\begin{itemize}
\item
  {\bf Lattice QCD (LQCD)}. It is a sophisticated numerical method with very demanding calculations lasting many years, even decades. But eventually in the case of $K\to\pi\pi$ decays and $K^0-\bar K^0$ mixing this method
  is expected to give the ultimate results for $\epe$, $\Delta I=1/2$ rule and
  $K^0-\bar K^0$ mixing, both in the SM and beyond it. For $K\to\pi\pi$  only
  results for the SM operators are known and they are far from being satisfactory. The
  ones for BSM $K^0-\bar K^0$ matrix elements are already known with respectable  precision and interesting results have been obtained for long distance
  contributions to $\Delta M_K$ \cite{Bai:2018mdv,Wang:2018csg}.
\item
  {\bf Dual QCD (DQCD)} proposed already in the 1980s \cite{Bardeen:1986vz}, significantly improved
  in the last decade \cite{Buras:2014maa} and also very recently as reported below. This approach, which opposed to LQCD is fully analytical, allows to obtain results for $K\to\pi\pi$ decays
  and $K^0-\bar K^0$ mixing much faster than it is possible with the LQCD,
typically within a few months. Therefore several relevant results have been obtained already in the 1980s and   confirmed within uncertainties by LQCD in the last decade. While not as accurate   as the expected ultimate LQCD calculations, it allowed already to calculate   hadronic matrix elements for all BSM operators entering $K\to\pi\pi$ decays   \cite{Aebischer:2018rrz}
and $K^0-\bar K^0$ mixing \cite{Buras:2018lgu}. Very importantly this approach allows to get the insight into the   QCD dynamics at low energy scales which is not possible using a purely   numerical method like LQCD. A good example
are four  hadronic matrix elements of BSM operators entering
$K^0-\bar K^0$ mixing  \cite{Buras:2018lgu}.   For a detailed
exposition of this approach see \cite{Buras:2014maa,Buras:2020xsm,Buras:2018hze}. More about it below.

\item
  {\bf Chiral Perturbation Theory (ChPT)} developed since 1978 \cite{Weinberg:1978kz,Gasser:1983yg,Gasser:1984gg,Ecker:1988te,Ecker:1994gg,Pich:1995bw,Cirigliano:2011ny} and discussed recently in \cite{Cirigliano:2019zjv,Cirigliano:2019ani} in the contex of $\epe$. It is based on global symmetries of QCD with the QCD dynamics
  parametrized by low-energy constants $L_i$ that enter the counter terms
  in meson loop calculations. $L_i$ can {be extracted
  from the data or calculated by LQCD. Presently this framework
  as applied to non-leptonic transitions has
  serious  difficulties in matching
  long distance (LD) and short distance (SD) contributions, a problem admitted
  by the authors of \cite{Cirigliano:2019ani,Gisbert:2020wkb}. For instance
  the expression for the matrix element of the QCDP operator $Q_6$  in terms of $L_5$ is only valid in the large $N$ limit, that is using factorization.
 On the other hand} the dominant QCD dynamics in
  Wilson coefficients is given by non-factorizable contributions.
  This problem is   absent in LQCD and DQCD as we will discuss below. 
  Therefore,   while the ChPT approach is very suitable for leptonic and semi-leptonic Kaon   decays, it can only provide partial information on $\epe$ and the $\Delta I=1/2$   rule in the form of isospin breaking effects and final state interactions (FSI). Yet, in the case of isospin breaking contributions to $\epe$ the difficulties  in the matching in question imply a rather significant error as we will see   below.
  \end{itemize}

This writing is arranged as follows. In Section~\ref{sec:2} I will briefly
describe the DQCD approach. In Section~\ref{sec:3} I will summarize my view on the present status of $\epe$ in the SM.  In Section~\ref{sec:4} I will
review the recent progress in the evaluation of $\epe$ within
the Weak Effective Theory (WET) and the Standard Model Effective Field Theory (SMEFT).  In Section~\ref{sec:4b} several comments on the dynamics behind
the $\Delta I=1/2$ rule in $K\to\pi\pi$ decays will be made.
A brief outlook in Section~\ref{sec:5} ends this presentation.

\section{ The Dual QCD Approach: A Grand View}\label{sec:2}
This analytic approach to $K\to\pi\pi$ decays and $K^0-\bar K^0$ mixing in   \cite{Bardeen:1986vz,Bardeen:1987vg,Buras:2014maa} is based on 
the ideas of 't Hooft and Witten \cite{'tHooft:1973jz,'tHooft:1974hx,Witten:1979kh,Treiman:1986ep} who studied QCD with a
 large number $N$ of colours. In this limit  QCD is dual to a  theory of weakly interacting mesons with the coupling $\ord(1/N)$ and in particular in the strict large $N$ limit it becomes a free theory 
of mesons, simplifying the calculations significantly. With  non-interacting mesons the factorization of matrix elements 
of four-quark operators  into matrix elements of quark currents and quark 
densities,  used adhoc in the 1970s and early 1980s, is automatic and can be considered as a property of QCD in this limit \cite{Buras:1985xv}. But the factorization cannot be the whole 
story as the most important QCD effects related to asymptotic freedom 
are related to non-factorizable contributions generated by exchanges of gluons.
In DQCD this role is played by meson loops that represent 
dominant non-factorizable contributions at the very low energy scales. Calculating these loops with a momentum cut-off $\Lambda$ one finds
then  that the factorization in question 
does not take place at values of $\mu \ge 1\gev$ at which Wilson coefficients 
are calculated, but  rather at very low  momentum transfer between colour-singlet currents or densities.  

Thus, even if  in the large $N$ limit the hadronic matrix elements factorize
and can easily be calculated, 
in order to combine them with the Wilson coefficients, loops in the meson 
theory have to be calculated. In contrast to chiral perturbation theory, in 
DQCD a physical cut-off $\Lambda$ is used in the integration over loop momenta.
 As discussed in detail in \cite{Bardeen:1986vz,Buras:2014maa} this allows
 to achieve  a much better matching with short distance contributions than it
is possible in ChPT, which uses dimensional regularization.  The cut-off $\Lambda$ is typically chosen around $0.7\gev$ when only pseudoscalar mesons are exchanged in the
loops \cite{Bardeen:1986vz} and can be increased up to $0.9\gev$ when contributions from lowest-lying vector mesons are taken into account as done in \cite{Buras:2014maa}. These calculations are done in a momentum scheme, but as 
demonstrated in \cite{Buras:2014maa}, they can be matched to the commonly 
used naive dimensional regularization (NDR) scheme. Once this is done it is justified to set $\Lambda\approx \mu$.

The application of DQCD to weak decays consists in any  NP model of the following steps:

{\bf Step 1:} At $\Lambda_\text{NP}$ one integrates out the heavy degrees of freedom and performs the RG evolution including Yukawa couplings and all gauge interactions present in the SM down to the electroweak scale. This evolution involves in addition to SM 
operators also beyond the SM (BSM) operators. This is the SMEFT.

{\bf Step 2:} At the electroweak scale $W$, $Z$, top quark and the Higgs are integrated out and the SMEFT is matched onto the WET with only SM quarks except the top-quark, the photon and the gluons. Subsequently QCD and QED evolution is performed down to scales $\ord(1\gev)$. The presence of many new operators
makes this evolution rather involved. Fortunately as we will report below
this evolution is known by now at NLO level in QCD.

{\bf Step 3:} Around scales of $\ord(1\gev)$ the matching to the theory of mesons
is performed and the quark evolution is followed by {\em the meson evolution} down to the factorization scale.

{\bf Step 4:} At the appropriate hadronic scale between the kaon and pion masses, the matrix elements of all operators are calculated in the 
large $N$ limit, that is using factorization of matrix elements into products of currents or densities.

I do not claim that these are all QCD effects responsible for non-leptonic 
transitions, but these evolutions based entirely on non-factorizable QCD
effects, both at short distance and long distance scales, appear to be the main 
bulk of QCD dynamics responsible for the $\Delta I=1/2$ rule, $\epe$ and $K^0-\bar K^0$ mixing. Past successes of this approach have been reviewed in 
\cite{Buras:2020xsm,Buras:2018hze,Buras:2018wmb}. They are related
in particular to the non-perturbative parameter $\hat B_K$ in $K^0-\bar K^0$
mixing and  $\Delta I=1/2$ rule \cite{Buras:2014maa}.
In fact DQCD allowed for the first time to identify already in 1986
the dominant mechanism behind this rule \cite{Bardeen:1986vz}. It is
simply the quark evolution from short distance scales, possibly involving NP,
down to scales of $\ord(1\gev)$, followed by meson evolution down to
the factorization scale around the kaon and pion masses.

In summary DQCD turns out to be an efficient approximate method for obtaining results for non-leptonic decays, years and even decades, before useful results from numerically sophisticated and demanding lattice calculations could be obtained. This will be particularly clear in Section~\ref{sec:4}, where we will turn
our attention to the results in the WET and the SMEFT which are to date
available only within DQCD approach.

\section{$\epe$ in the SM}\label{sec:3}

The situation of $\epe$ as of March 2022 can be summarized as follows.

 The   experimental world average from NA48 \cite{Batley:2002gn} and
  KTeV \cite{AlaviHarati:2002ye, Worcester:2009qt} collaborations reads
  \be
    \label{EXP}
    \boxed{(\epe)_\text{exp} 
    = (16.6 \pm 2.3) \times 10^{-4}\,.}
  \ee

The most recent result from LQCD, that is from 
the RBC-UKQCD collaboration
\cite{Abbott:2020hxn}, reads
\be
  \label{RBCUKQCD}
  \boxed{(\epe)_{\rm SM} 
  = (21.7 \pm 8.4) \times 10^{-4} \,,\qquad (\text{RBC-UKQCD}-2020),}
\ee
where statistical, parametric and systematic uncertainties have been added in
quadrature. The central value is by an order of magnitude larger than
the central 2015 value presented by this collaboration but is subject
to large systematic uncertainties which dominate the quoted error. It 
is based on the improved values of the hadronic matrix elements of QCDP,
includes the Wilson coefficients at the NLO level but does not
include  isospin breaking effects, charm contributions and NNLO QCD effects.

The most recent 
estimate of $\epe$ in the SM from ChPT \cite{Cirigliano:2019ani,Gisbert:2020wkb} reads
  \be
  \label{Pich}
  \boxed{(\epe)_\text{SM}   = (14 \pm 5) \times 10^{-4} \,,\qquad (\text{ChPT}-2019).}
\ee
  The large error is related to the problematic matching
  of LD and SD contributions in this approach which can be traced back
  to the absence of meson evolution in this approach. Here the isospin breaking corrections are included
  but as discussed below they are subject to significant uncertainties
  again related to the problematic matching   of LD and SD contributions.

  Finally, based on the insight from DQCD obtained in collaboration
  with Jean-Marc G{\'e}rard  \cite{Buras:2015xba,Buras:2016fys} one finds
  \cite{Buras:2020wyv}
  \be\label{AJBFINAL}
 \boxed{(\epe)_{\rm SM}= (5\pm2)\cdot 10^{-4},\qquad (\text{DQCD}-2020).}
 \ee

 While the results in (\ref{RBCUKQCD}) and (\ref{Pich}) are fully consistent
 with the data in (\ref{EXP}), the DQCD result in (\ref{AJBFINAL}) if confirmed
 by other groups would one day imply a significant anomaly and NP at work.

 Let us then have a look at the phenomenological  expression for $\epe$ in the
 SM in order to understand the origin of these differences.
This  formula  
 presented in \cite{Aebischer:2020jto} reads
\begin{equation}
\left(\frac{\varepsilon'}{\varepsilon}\right)_{\text{SM}} =  
\IM\lambda_{\rm t}\cdot \left[\,
\big(1-\hat\Omega_{\rm eff}\big) \big(-2.9 + 15.4\,\bsi(\mu^*)\big) + 2.0 -8.0\,\bei(\mu^*) \,\right].
\label{AN2020}
\end{equation}
It includes NLO QCD corrections to the QCD penguin  (QCDP) contributions and NNLO contributions to electroweak penguins (EWP). The coefficients in this formula
and the parameters $\bsi$ and $\bei$, conventionally normalized to unity at the factorization scale, are scale dependent. Their values for
different scales are collected in Table~1 of \cite{Aebischer:2020jto}. Here
we will set  $\mu^*=1\gev$ because at this scale it is most convenient to compare
the values for $\bsi$ and $\bei$ obtained in the three  non-perturbative approaches in question.
The four contributions in (\ref{AN2020}) are dominated
by the  following operators:
\begin{itemize}
\item
  The terms involving the non-perturbative parameters $\bsi$  and $\bei$ 
contain only the contributions from the dominant QCDP  operator
$Q_6$ and the dominant EWP operator $Q_8$, respectively. There are two main reasons
why $Q_8$ can compete with $Q_6$ here despite the smallness of the electroweak couplings relative to the QCD one. In the basic formula for $\epe$ its contribution is 
enhanced relative to the $Q_6$'s one by the  factor ${\RE A_0/\RE A_2}=22.4$ with  $A_{0,2}$ being isospin amplitudes. In addition its Wilson coefficient is enhanced for the large top-quark mass which is not the case of the $Q_6$'s one
 \cite{Flynn:1988ve,Buchalla:1989we}. The expressions for
these two operators and the remaining operators mentioned below are well known
and can be found in \cite{Buras:2020wyv,Buras:2020xsm}.
\item
The term $-2.9$ is fully dominated by the QCDP operator $Q_4$.
\item
  The term $+2.0$ is fully dominated by EWP operators $Q_9$ and $Q_{10}$.
\item
  The quantity $\hat\Omega_{\rm eff}$ represents the isospin breaking corrections
  and QED corrections beyond EWP contributions. It is not in the ballpark of a few percent  as one
  would naively expect,  because in $\epe$ it is again
enhanced by the  factor ${\RE A_0/\RE A_2}=22.4$.
\item
${\IM\lambda_{\rm t}}$ is the CKM factor that within a few percent is in the ballpark of ${1.45\cdot 10^{-4}}$.
\end{itemize}

Now at $\mu=1\gev$ the values of $\bsi$ and $\bei$ in the three non-perturbative
approaches are found to be:
 \begin{equation}\label{LATB8}
   \bsi(1\gev)   = 1.49 \pm 0.25,  \quad
  \bei(1\gev) 
    = 0.85 \pm 0.05\,,\quad  (\text{RBC-UKQCD}-2020).
 \end{equation}

 \begin{equation}\label{ChPTB}
   \bsi(1\gev)   = 1.35 \pm 0.20,  \quad
  \bei(1\gev) 
    = 0.55 \pm 0.20\,,\quad  (\text{ChPT}-2019).
 \end{equation}

 \begin{equation}\label{DQCDB}
   \bsi(1\gev)   \le 0.6,  \quad
  \bei(1\gev) 
    = 0.80 \pm 0.10\,,\quad  (\text{DQCD}-2015).
 \end{equation}

Let us begin  with the good news. There is
a very good agreement between
LQCD and DQCD as far as EWP contribution to $\epe$ is concerned. This implies
that this contribution to $\epe$, that is unaffected by leading isospin breaking corrections, is already
known within the SM with acceptable accuracy:
\be
  \label{EWPSM}
  \boxed{(\epe)^{\text{EWP}}_\text{SM}   = - (7 \pm 1) \times 10^{-4} \,,\qquad (\text{LQCD~and~DQCD}).}
\ee
Because both LQCD and DQCD can perform much better in the case of EWPs than in the case of QCDPs I expect that this result will remain with us for the coming years.

On the other hand the value from ChPT $\bei\approx 0.55$ \cite{Cirigliano:2019ani} implies on the basis of the last two terms
in (\ref{AN2020}) the EWP contribution  roughly by a factor of 2 below the result in (\ref{EWPSM}). This lower value originates in the 
suppression caused by FSI which seems to be not supported
by the results from the RBC-UKQCD collaboration. Yet, the uncertainty
in the ChPT estimate is large. It would be good to clarify this difference in the
future.

The case of QCDPs is a different story. Here the LQCD value of $\bsi$  overshoots the DQCD
one by more than a factor of two 
and consequently despite the agreement on EWP contribution the result
in (\ref{RBCUKQCD}) from  RBC-UKQCD differs by roughly a factor of four
from the one of DQCD in  (\ref{AJBFINAL}).

This difference will surely  decrease when the RBC-UKQCD collaboration
will include isospin breaking and QED corrections. As far as these corrections
are concerned LQCD, ChPT, DQCD use in their estimates respectively  
\be
\hat\Omega_{\rm eff}=0,\qquad
\hat\Omega_{\rm eff}^{(8)}=(17\pm9)\, 10^{-2},\qquad
\hat\Omega_{\rm eff}^{(9)}=(29\pm7)\, 10^{-2},
\ee
  where the index ``(9)'' indicates that the full nonet of pseudoscalars
  has been taken into account in the case of DQCD. This means that
  also $\eta-\eta^\prime$ mixing has been taken explicitly into account  \cite{Buras:2020pjp}. In the octet scheme, necessarily used in ChPT, this mixing  is buried in a poorly   known low-energy constant $L_7$ and 
    I expect that including the
  effect of this mixing in $\epe$ will remain a challenge for ChPT for 
  some time. Yet, it is evident from the analysis in \cite{Buras:2020pjp} that 
  this mixing enhances significantly $\hat\Omega_{\rm eff}$, the fact known already for 35 years \cite{Donoghue:1986nm,Buras:1987wc}.
  
  There is still another  reason why the DQCD result in (\ref{AJBFINAL}) is much
  lower than the ones in LQCD and ChPT. We include NNLO QCD corrections to
  both QCDP and EWP contributions that together provide a downward shift
  of $\epe$ in the ballpark of $3\times 10^{-4}$. While ChPT experts
  criticized us for ignoring FSI, this is really a fake news. We devoted
  a full paper to FSI \cite{Buras:2016fys}. While we agreed that in DQCD precise estimate of
  FSI is difficult, we gave arguments why the effect of the enhancement of
  $\bsi$ by FSI is likely to be smaller than its suppression by the meson evolution.   In fact FSI are included in (\ref{AJBFINAL}) with $\bsi\approx 1.0$
  corresponding to the upper limit when subleading FSI would fully cancel the
  suppression by the leading meson evolution.

  In Table~\ref{tab:SUP} I summarize the main SM dynamics which in my view is responsible   for the strong suppression of $\epe$ below the experimental value. The question marks in the case of ChPT mean that presently it is not clear
  how well the low energy constants $L_5$ and $L_7$ will describe meson evolution and $\eta-\eta^\prime$ mixing, respectively. To be fair one should add that
  while RBC-UKQCD and ChPT experts claim  to have a satisfactory treatment of FSI for penguin operators, in
  the case of DQCD it can only be argued at present that the subleading FSI can at best
  cancel the suppression of QCDP contribution by the leading meson evolution
  \cite{Buras:2016fys}.

  Table~\ref{tab:SUP} summarizes my view of the controversy between the three approaches. In other words my message for unbiased non-experts to take home is the following
  \begin{itemize}
  \item
    RBC-UKQCD collaboration and ChPT experts do not claim that there is no NP in
    $\epe$. But as of March 2022 their methods are not sufficiently powerful to see the anomaly in $\epe$. I expect that RBC-UKQCD collaboration will see
    this anomaly after they reached their goals for coming years summarized recently in \cite{Blum:2022wsz}.
  \item
    DQCD approach, even if approximate, can much faster see the underlying dynamics responsible for possible anomaly in $\epe$ because it is an  analytical approach
    and important QCD dynamics is not hidden in the numerical values of
    hadronic matrix elements evaluated by LQCD nor hidden  in low energy constants
    $L_i$ of ChPT.
  \end{itemize}

  I am truly delighted that at least one group strongly believes in NP in $\epe$
  at work. How boring would be a situation of $\epe$ if also DQCD found $\epe$
  the ballpark of the values presented by RBC-UKQCD and ChPT. Simultaneously
  I understand that both groups are, in contrast to me,  not motivated to
  study NP in $\epe$. I wish them therefore luck in  improving their calculations   and will now go beyond the SM.

 \begin{table}[!tb]
\centering
\begin{tabular}{|c|c|c|c|}
\hline
 & DQCD  & RBC-UKQCD & ChPT  \\
\hline
Large $m_t$ &  $\star$   &  $\star$ &  $\star$ \\
Meson evolution &  $\star$   &  $\star$ &  $L_5(?)$ \\
QCDP at NNLO &  $\star$   &  $-$ &  $-$ \\
EWP at NNLO &  $\star$   &  $-$ &  $-$ \\
IB (Octet) &  $\star$   &  $-$ &  $\star$ \\
$\eta-\eta^\prime$ mixing &  $\star$   &  $-$ &  $L_7(?)$ \\
\hline
\end{tabular}
\caption{\it Comparision of various suppression mechanism of $\epe$ in the SM taken ($\star$) or not taken ($-$) into account in a given approach. See text for more details.
}\label{tab:SUP}
\end{table}

 \section{$\epe$ beyond  the SM}\label{sec:4}

 A new era for $\epe$ began in 2015, when the RBC-UKQCD collaboration  \cite{Bai:2015nea,Blum:2015ywa} presented their first results for $\epe$ with
 the central value by rougly a factor of 15 below their 2020 value in
 (\ref{RBCUKQCD}).  This inspired many authors to search for  NP which would explain this anomaly. I will not review
 these analyzes again as I have done it already in \cite{Buras:2020wyv,Buras:2020xsm,Buras:2019vik}. But I would like to stress that all these analyses with a few exceptions involved only the SM operators, that is the first term in
 (\ref{Kpipi}) in wich NP modifies the Wilson coefficients of the SM operators,
 in particular the one of $Q_8$.

 Here I would like to confine first the discussion to the second term in
 (\ref{Kpipi}) and subsequently to $\epe$ in the framework of the WET and the SMEFT  which include all operators contributing to $\epe$ in these effective theories.

 In fact since 2018 a significant progress   towards the general search for
 NP in $\epe$ with the help of DQCD and more generally has been made. I will just list all papers written by us in this context and describe in more details only the most recent ones. Here we go.

\begin{itemize}
\item
The first to date results for the $K\to\pi\pi$ matrix elements of 
the chromo-magnetic dipole operators \cite{Buras:2018evv} that
are compatible with the LQCD results for $K\to\pi$ matrix elements
of these operators obtained earlier in \cite{Constantinou:2017sgv}.
\item
The first to date calculation of $K\to\pi\pi$ matrix elements of {\em all} four-quark BSM operators,
including scalar and tensor operators, by DQCD \cite{Aebischer:2018rrz}.
\item
The derivation of a master formula for $\epe$ \cite{Aebischer:2018quc}, which
can be applied to any theory beyond the SM in which the
Wilson coefficients of all contributing operators have been calculated at the
electroweak scale. The relevant hadronic matrix elements of BSM operators used 
in this formula are
from the DQCD, as lattice QCD did not calculate them yet, 
and the SM ones from LQCD.
\item
This allowed to perform the first to date  model-independent anatomy of the ratio $\epe$
in the context of  the $\Delta S = 1$ effective theory with operators invariant
under QCD and QED and in the context of the SMEFT with the operators invariant under the full SM gauge group \cite{Aebischer:2018csl}.
\item
  Finally the insight from DQCD \cite{Buras:2018lgu} into the values of BSM $K^0-\bar K^0$ elements obtained by LQCD  made sure that the meson evolution is
  hidden in lattice calculations.
\end{itemize}
  The main messages from these papers are as follows:
\begin{itemize}
\item
  The inclusion of the meson evolution in the phenomenology of any non-leptonic
  transition like $K^0-\bar K^0$ mixing and $K\to\pi\pi$ decays with $\epe$ and
  the $\Delta I=1/2$ rule is mandatory!
\item
  Meson evolution is hidden in LQCD results, but among analytic approaches only
  DQCD takes this important QCD dynamics into account. Whether meson evolution
  is present in the low energy constant $L_5$ of ChPT is an interesting question, still to be answered.
\item
  Most importantly, the meson evolution  
turns out to have the pattern of operator mixing, both for SM and BSM operators,
to agree with the one found perturbatively at short distance scales. This allows for a satisfactory, even if approximate, matching between Wilson coefficients and hadronic matrix elements.
\end{itemize}
  
The BSM part of $\epe$ in our 2018 papers was performed at LO in QCD RG
quark evolution. In the more recent papers we made progress by including
NLO QCD corrections and in this context we derived master formulae not only
for $\epe$ and $K\to\pi\pi$ but generally for non-leptonic $\Delta F=1$ and $\Delta F=2$ transitions. Let me briefly summarize these papers.

In \cite{Aebischer:2021raf} we reconsidered the complete set of four-quark operators in the
WET for non-leptonic $\Delta F=1$ decays that
govern $s\to d$ and $b\to d, s$ transitions in
the SM and beyond, at NLO in QCD. In particular we discussed  the issue of
transformations between operator bases beyond leading order to
facilitate the matching to high-energy completions or the SMEFT
at the electroweak scale.
As a first step towards a SMEFT NLO analysis of $K\to\pi\pi$ and
non-leptonic $B$-meson decays, we calculated the relevant WET Wilson
coefficients including two-loop contributions to their renormalization
group running, and expressed them in terms of the Wilson coefficients in
a particular operator basis, the so-called JMS basis \cite{Jenkins:2017jig} for which the one-loop matching to SMEFT
is already known \cite{Dekens:2019ept}.

In \cite{Aebischer:2021hws} we have
 presented for the first time the NLO master formula for the BSM part of
$\epe$  expressed in terms of the Wilson coefficients of all contributing
 operators of WET evaluated at the electroweak scale. To this end we use  again
 JMS basis and the results from  \cite{Aebischer:2021raf}.
The relevant hadronic matrix elements of BSM operators at the electroweak
scale are taken from Dual QCD approach and the SM ones from lattice QCD.
It includes the renormalization group evolution and quark-flavour threshold
effects at NLO in QCD from hadronic scales, at which these matrix elements
have been calculated, to the electroweak scale.

This updated master formula for the BSM part of $\epe$ reads
reads \cite{Aebischer:2021hws}
\begin{align}
  \label{eq:master}
  \left(\frac{\varepsilon'}{\varepsilon}\right)_\text{BSM} &
  = \;\; \sum_b  P_b(\muEW)
    \im \big[ C_b(\muEW) - C^\prime_b(\muEW) \big]
    \times (1 \TeV)^2,
\end{align}
with the sum over $b$ running over the Wilson coefficients $C_b$ of all
operators in the JMS basis and their chirality-flipped counterparts
denoted by $C_b'$. The relative minus sign accounts for the fact that
their $K\to\pi\pi$ matrix elements differ by a sign. Among the contributing
operators are also operators present already in the SM, but their WCs in \eqref{eq:master} are meant to include only BSM contributions. The numerical values
of $P_b(\muEW)$ are collected in \cite{Aebischer:2021hws}. Calculating
$C_b(\muEW)$ and $C^\prime_b(\muEW)$ in a given BSM scenario this master
formula allows in no time to calculate the BSM contribution to $\epe$.

Another master formula, this time for $\Delta F=2$ transitions in the SMEFT
has been presented in \cite{Aebischer:2020dsw}. We illustrated it on a number of simplified
models containing colourless heavy gauge bosons ($Z^\prime$) and scalars
and models with coloured heavy gauge bosons ($G_a^\prime$) and scalars.
Also the cases of vector-like quarks and leptoquarks are briefly considered.
See brief summary of this work in \cite{Aebischer:2022ppu}.

The next important step is the inclusion of NLO QCD corrections to the RG evolution
in the SMEFT. The important benefit of these new contributions is that they allow to remove renormalization scheme dependences present both in the one-loop
  matchings between the WET and SMEFT and also between SMEFT and a chosen UV
  completion. For $\Delta F=2$ transitions this step has been already made very
  recently in \cite{Aebischer:2022anv}.

\section{Comments on the $\Delta I=1/2$ Rule}\label{sec:4b}
I have described the history of this rule, the ratio 
 ${\RE A_0/\RE A_2}=22.4$,
in Section 7 of \cite{Buras:2020wyv} and 
in particular in Section 7.2.3 of my recent book \cite{Buras:2020xsm}.
From this it is evident that the credit for the identification of the
 basic dynamics behind this rule should go to the authors of \cite{Bardeen:1986vz} who demonstrated that the current-current operators and not QCDP operators
 are dominantly responsible for this rule. This has been  confirmed more than 30 years later
 by the RBC-UKQCD collaboration \cite{Boyle:2012ys,Abbott:2020hxn} basically ignoring our work.

 Leaving this issue here aside what is really puzzling for me is there interpretation of the dynamics behind this rule as they state from first principles. According to them
 this rule originates in the relation between two leading contractions
\be\label{CON2}
\boxed{2}=-K~\boxed{1} 
\ee
induced by QCD. The factor $K$ depends on the scale at which
these contractions are evaluated. In 2012 they were evaluated
at $\mu=2.15\gev$, in 2020 at  $\mu=4.0\gev$. 

In order to make my point let us follow the exercise performed by us in
Section 9 of \cite{Buras:2014maa} and extract the values of these contractions from the experimental values 
 for ${\rm Re}A_0$ and  ${\rm Re}A_2$
\cite{Zyla:2020zbs}:
\be\label{N1}
{\rm Re}A_0= 27.04(1)\times 10^{-8}~\gev,
\quad
\quad {\rm Re}A_2= 1.210(2)   \times 10^{-8}~\gev.
\ee
Adjusting the expressions for the isospin amplitudes in 
\cite{Boyle:2012ys} to our normalization we have
\be\label{L1}
({\rm Re}A_0)=\frac{G_F}{\sqrt{2}}V_{ud}V_{us}^*\left(\frac{\sqrt{2}}{3}\right)
\left[z_1\left(2~\boxed{2}-\boxed{1}\right)+z_2\left(2~\boxed{1}-\boxed{2}\right)\right],
\ee

\be\label{L2}
({\rm Re}A_2)=\frac{G_F}{\sqrt{2}}V_{ud}V_{us}^*(\frac{2}{3})
(z_1+z_2)\left(\boxed{1}+\boxed{2}\right)~,
\ee
where $z_{1,2}$ are the real parts of the Wilson coefficients of the current-current operators $Q_{1,2}$.
 With the  NDR-${\rm \overline{MS}}$ 
 values $z_1=-0.199$ and $z_2=1.090$ at  $\mu=4.0\gev$,
the data in (\ref{N1}) can be reproduced with 
\be\label{CON1}
\boxed{1}=0.0865\gev^3,\quad        \boxed{2}= -0.0752 \gev^3,\quad 
\boxed{2}=-0.87~ \boxed{1}.
\ee
Thus at  $\mu=4.0\gev$ we need $K=0.87$ to reproduce the data. In 2012 RBC-UKQCD found
$K\approx 0.7$ at  $\mu=2.15\gev$ which was not sufficient to reproduce
the data but in 2020 their value of $K$ is in the ballpark of $0.85$ with
a significant error. Certainly it is an important result.

So far so good. Now using this relation in (\ref{L2}) the RBC-UKQCD
finds strong suppression of ${\rm Re}A_2$ and concludes that this suppression
is the main origin of the $\Delta I=1/2$ rule.

 This is a surprizing interpretation because until the 2012
 RBC-UKQCD paper \cite{Boyle:2012ys}, the particle community thought that it is the strong enhancement of $\RE A_0$ and some moderate
 suppression of  $\RE A_2$  responsible for this rule.

 This very original interpretation from first principles, also repeated at this workshop, differs totally from the one of DQCD as presented in our
 papers, in particular in \cite{Bardeen:1986vz,Buras:2014maa} but also recently in \cite{Buras:2020wyv,Buras:2020xsm}. I do not want to repeat this interpretation again because it is easy to demonstrate in a few lines without using super computers or even calculating meson loops that the RBC-UKQCD interpretation 
 is incorrect.

 The point is that in order to find out the impact of the  QCD dynamics on these two amplitudes, that is responsible for this rule, one has to ask first the question what would be
 the values of these two amplitudes and of the corresponding contractions in the absence of QCD.
 Of course we cannot fully switch off QCD because we want to have quarks confined in mesons but we can switch off non-factorizable contributions both in SD and LD by going to the strict large $N$ limit. In this limit only the
 operator $Q_2$  contributes, its coefficient is $z_2=1$ and its matrix element
can be calculated precisely using factorization valid in QCD in this limit.
One finds then\footnote{see \cite{Buras:2014maa,Buras:2020xsm} for a detailed presentation.}:
\be\label{LO}
{\rm Re}A_0=3.59\times 10^{-8}\gev ,\quad   {\rm Re}A_2= 2.54\times 10^{-8}\gev~, \quad \frac{\RE A_0}{\RE A_2}=\sqrt{2},
\ee
in plain disagreement with the experimental value of 22.4.
It should be emphasized that the explanation of the  missing enhancement factor of $15.8$ through some dynamics must simultaneously give the correct values
for the experimental values of ${\rm Re}A_0$ and  ${\rm Re}A_2$ in (\ref{N1}).
This means that these dynamics should suppress  ${\rm Re}A_2$ by a factor of $2.1$, not more, and enhance ${\rm Re}A_0$ by a factor of $7.5$. This tells us 
that while the suppression of  ${\rm Re}A_2$  is an important ingredient in 
the $\Delta I=1/2$ rule, it is $\underline{not}$ the main origin of this rule.
 It is the enhancement of  ${\rm Re}A_0$,  as
already emphasized in \cite{Shifman:1975tn}, even if, in contrast to this paper, as demonstrated in \cite{Bardeen:1986vz}, the
current-current operators are responsible dominantly for this rule and not 
QCD penguins.

One can also do this exercise using contractions to find in the factorization
limit \cite{Buras:2014maa}
\be\label{LargeN}
\boxed{1}=0.0211~\gev^3, \qquad \boxed{2}=0, \qquad (\mu\approx 0)
\ee
which drastically differ from the values of contractions in (\ref{CON1}).
In other words the main impact of QCD is to change the values of contractions
from (\ref{LargeN}) to the ones in the ballpark of the ones in (\ref{CON1})
and this change have significantly larger effect on ${\rm Re}A_0$ than on
${\rm Re}A_2$.

The simple discussion above shows that the RBC-UKQCD interpretation of the dynamics behind $\Delta I=1/2$ rule cannot be correct. But  working numerically
at $4\gev$ and being not able to switch-off non-factorizable QCD interactions in LQCD one simply  cannot see properly what is really going on. Yet, their numerical result is very important and one should
congratulate them for it. In particular because their central value for the ratio in question, $19.9\pm5.0$,  is closer to the data than present one from DQCD, $16\pm2$  \cite{Buras:2014maa}.
In the latter approach an additional enhancement could come from FSI which
this time are not compensated by meson evolution as was the case of $\epe$.
It could also come from NP \cite{Buras:2014sba}.

Let me then repeat.
The dominant dynamics behind the $\Delta I=1/2$ rule for $K\to\pi\pi$ isospin amplitudes is our beloved QCD as also
confirmed by RBC-UKQCD. But the correct physical interpretation differs drastically
from the one provided by them. It is simply the
    {\em quark evolution} from $M_W$ down to scale $\ord(1\gev)$ as analysed first     by  Altarelli and Maiani \cite{Altarelli:1974exa} and Gaillard and Lee \cite{Gaillard:1974nj}, followed by the  {\em meson evolution} down to very low scales at which  QCD becomes  a theory of weakly interacting mesons and free theory of mesons in the strict large $N$ limit \cite{'tHooft:1973jz,'tHooft:1974hx,Witten:1979kh,Treiman:1986ep}. This non-perturbative evolution 
    within  the Dual QCD approach dominates by far the enhancement of
    ${\RE A_0}$ over ${\RE A_2}$ as demonstrated by Bardeen, G{\'e}rard and myself in  \cite{Bardeen:1986vz,Buras:2014maa}. The relation of contractions
    in (\ref{CON1}) cannot provide this interpretation because at
    $\mu=4\gev$ the computer only tells us what the values of these
    contractions are.

\section{Outlook}\label{sec:5}

Let me finish this presentation by stating what should be improved 
in the three non-perturbative approaches  in order to clarify the situation of $\epe$ in the SM.

{\bf RBC-UKQCD:} 
Here a very important progress would be the calculation of isospin breaking and QED corrections including also $\eta-\eta^\prime$ mixing. Another important issue
are charm contributions still missing in RBC-UKQCD calculations. The prospects
for improvements in the coming years are good \cite{Blum:2022wsz}.

{\bf ChPT:} 
First of all matching to short distance which would amount, as far as I understand,  at least to the determination of $L_5$. Better inclusion of  $\eta-\eta^\prime$ mixing with the help of $L_7$.

{\bf DQCD:} 
Here certainly one should have a look at the subleading FSI.

It would also be very important if other LQCD groups contributed to these investigations and in particular several LQCD groups calculated BSM hadronic
matrix elements. Here some promising results are signalled by two LQCD groups
\cite{Ishizuka:2018qbn,Donini:2020qfu}.

However, in order to identify possible NP also correlations of $\epe$ with other processes, in particular rare Kaon decays,
have to be studied as summarized recently in \cite{Aebischer:2022vky}.
Definitely there are exciting times ahead of us!

\section*{ Acknowledgements}
 
{I would like to thank  Jean-Marc G{\'e}rard for invaluable   comments on the manuscript and him,   Jason Aebischer, Christoph Bobeth, 
 Jacky Kumar, Mikolaj Misiak and David Straub } for exciting time we spent
together 
analyzing the topics discussed in this talk.
I would like to thank the organizers of WG3   for inviting me to present these results.
The research presented in this report was dominantly financed and done in the context of the ERC Advanced Grant project ``FLAVOUR'' (267104).  It was also partially supported by the 
DFG cluster of excellence ``Origin and Structure of the Universe''.

\bibliographystyle{JHEP}
\bibliography{Bookallrefs}
\end{document}